\documentclass[apjl]{emulateapj}
\usepackage{apjfonts}
\slugcomment{To appear in {\it The Astrophysical Journal Letters.}}
\lefthead{Tsuribe \& Omukai}
\righthead{Fragmentation of Low-metallicity Clouds}

\begin{document}

\title{
Dust-cooling--induced Fragmentation of Low-metallicity Clouds
}

\author{Toru Tsuribe\altaffilmark{1} and Kazuyuki Omukai\altaffilmark{2}}

\altaffiltext{1}{Osaka University, Toyonaka, Osaka 540-0053, Japan; 
tsuribe@vega.ess.sci.osaka-u.ac.jp }

\altaffiltext{2}{
National Astronomical Observatory of Japan, Mitaka, Tokyo 181-8588, Japan; 
omukai@yso.mtk.nao.ac.jp }


\begin{abstract}
Dynamical collapse and fragmentation of low-metallicity cloud cores  
is studied using three-dimensional hydrodynamical calculations, with
particular attention devoted whether the cores fragment in the 
dust-cooling phase or not.
The cores become elongated in this phase, being unstable 
to non-spherical perturbation due to the sudden 
temperature decrease. 
In the metallicity range of $10^{-6}-10^{-5}Z_{\odot}$, cores 
with an initial axis ratio $\ga 2$ reach a critical 
value of the axis ratio ($\ga 30$) and fragment into 
multiple small clumps.
This provides a possible mechanism to produce low-mass stars 
in ultra-metal-poor environments.
\end{abstract}


\section{Introduction}
It has been suggested that the first stars in the universe were very massive, 
typically $\ga 100M_{\sun}$ (e.g., Abel, Bryan, \& Norman 2000, 
2002; Bromm et al. 1999, 2002; Omukai \& Palla 2001, 2003).
On the other hand, in present-day local star-forming regions, 
typical stars are low-mass objects $\la 1M_{\sun}$.
One of the major factors distinguishing these two modes of star formation 
is the degree of metal enrichment. 
At some point in the history of the universe, metallicity exceeds a 
critical value and the transition to low-mass star formation mode 
is thought to take place.
Since the exact value of critical metallicity is crucial in modeling 
such important problems as galaxy formation and  
reionization of the intergalactic medium etc., 
some authors have tried to establish this value.
Bromm, Ferrara, Coppi \& Larson (2001) studied the fragmentation of 
low-metallicity clouds, and concluded that clouds with metallicity 
$\ga 5\times 10^{-4}Z_{\sun}$ fragment into smaller pieces owing to 
the fine-structure line cooling of carbon and oxygen.
Taking into account possible variations in the elemental abundance 
ratio, Bromm \& Loeb (2003) derived individual critical abundances 
of C$^{+}$ and O.
Santoro \& Shull (2005) included silicon and iron in a similar analysis.  
However, due to limitations imposed by inclusion of only the atomic 
line cooling, these authors only studied the thermal evolution 
in the low-density regime. 
The mass scale of such fragments is 
still high $\ga 10-100M_{\sun}$, and another phase of fragmentation 
is necessary to produce low-mass ($\la 1M_{\sun}$) objects.
Regarding this problem, based on the thermal evolution described 
by Omukai (2000), 
Schneider et al. (2002, 2003) pointed out that another episode of 
fragmentation can be caused by dust cooling at higher densities 
(say, $\ga 10^{10} {\rm cm^{-3}}$).
They concluded that dust-induced fragmentation can take place 
even with metallicity of $10^{-5\pm 1} Z_{\sun}$ and this produces 
low-mass ($\la 1M_{\sun}$) fragments.
However, their treatment of dynamical evolution was too simple and 
their prediction of fragmentation remained speculative.

In the previous paper (Omukai et al. 2005), using an updated 
thermal evolution and the linear theory for growth of elongation 
(Hanawa \& Matsumoto 2000), we evaluated fragmentation mass scale
under the assumption that fragmentation takes place when 
the elongation exceeds a critical value ($\sim 1$) in the linear theory.
However, an approximately spherical core, for which the linear theory is
applicable, is not subject to fragmentation. Only a highly elongated, 
filamentary (or disk-like) object is gravitationally unstable and 
easily fragments (e.g., Larson 1995; Tsuribe \& Inutsuka 1999b).
Consequently, a discussion based on linear theory needs to be justified
by numerical calculations.

In consideration of the above, here we follow the evolution of 
low-metallicity cores during the dust-cooling phase using three-dimensional 
hydrodynamical calculations and discuss 
the conditions for fragmentation into low-mass clumps. 
Our results confirm that the previous analysis by Omukai et al. (2005) 
is broadly correct: fragmentation does occur in the dust-cooling phase 
and sub-solar mass fragments are indeed produced.

\section{Numerical Modeling}

H$_2$-line cooling induces fragmentation of 
low-metallicity ($\la 10^{-4}Z_{\sun}$) interstellar clouds  
around $10^{4} {\rm cm^{-3}}$, and cloud cores are produced.
Subsequently, the cores start a so-called run-away collapse, 
where they contract at almost the free-fall rate. 
Here we study evolution of such cores using 
three-dimensional hydrodynamical calculation. 
Since the cores are already self-gravitating, the dark-matter gravity 
is not considered.
For simplicity, the cores are assumed to be non-rotating and isolated. 

During the run-away collapse phase, a metal-free pre-stellar core 
obeys the self-similar flow with $\gamma \simeq 1.1$, 
as indicated by one-dimensional radiation
hydrodynamical calculations (Omukai \& Nishi 1998).
Here we choose such a self-similar solution as an unperturbed 
initial state.
The initial central number density is $10^{10}{\rm cm^{-3}}$ where 
the dust cooling is about to become effective.

We add both non-spherical density/velocity and
random-velocity perturbations to the initial state.
As a non-spherical density perturbation, the sphere is elongated to a 
prolate with expanding in the $z$-direction and shrinking in $x$- and 
$y$- directions homologously while the central density is kept constant. 
We adopt prolate elongation because we find that the initial oblateness 
leads the amplitude of elongation to saturation in the non-linear 
regime before fragmentation takes place.
Elongation of the core is defined by 
${\cal{E}}_0 = a/b - 1$,   
where $a$ ($b$) is the short (long, respectively) core axis length.
Here, cases with an initial elongation of
${\cal{E}}_0 = 0.1, 0.32, 0.56$, and $1.0$ are considered.
Non-spherical velocity perturbation is added to the 
unperturbed flow. 
The amplitude of non-spherical velocity perturbation is assumed 
to be proportional to the displacement from the sphere and chosen so 
that the correct linear growth (or damping) rate is achieved.
Additionally, a random velocity field with an amplitude of $0.1c_s$ 
is added, where $c_s$ is the isothermal sound velocity at the center.

Solving full thermal equations along with hydrodynamics 
is numerically too expensive.
Here we take the following simplified approach:
First, using a one-zone model, we solve thermal and chemical processes 
under the assumption of free-fall collapse.
The model is the same as in Omukai et al. (2005), except the dust model, 
for which we use dust produced by the pair-instability supernova of a 
metal-free $195M_{\sun}$ star 
(Schneider, Salvaterra, \& Ferrara 2004; Schneider et al. 2006).
The effective equations of state calculated with this model 
are shown in Figure 1 for metallicities $10^{-6}, 10^{-5}$, 
and $10^{-4}Z_{\sun}$. 
Next, for easier analysis and interpretation of the result, 
we approximate the above equations of state using
the following barotropic relations:
the ratio of specific heat 
$\gamma \equiv d{\rm log} p/d {\rm log} \rho$ 
is given by 
\begin{equation}
\label{eq:eos}
\gamma = \left\{
\begin{array}{ll}
1.1 & \mbox{$(n < n_{1})$} \\
0.5 & \mbox{$(n_{1} < n< n_{2})$} \\
1.1 & \mbox{$(n_{2} < n < n_{3})$} \\
1.4 & \mbox{$(n_{3} < n)$}, 
\end{array}
\right.
\end{equation}
where ${\log}~n_{1}=6-{\rm log}~Z/Z_{\sun}$, 
${\log}~n_{2}=11.1-0.4 {\rm log}~Z/Z_{\sun}$, and
${\log}~n_{3}=8.47-1.2 {\rm log}~Z/Z_{\sun}$.
In Figure 1, these relations are also shown by dotted lines. 
The first phase ($n < n_{1}$) corresponds to the H$_2$-cooling phase, 
the second ($n_{1}<n<n_{2}$) and third ($n_{2}<n<n_{3}$) to 
the dust-cooling phase before and after the gas-dust coupling, 
and the last ($n_{3}<n$) to the adiabatic contraction 
after the core becomes optically thick at $n_{3}$.

\begin{figure}
\plotone{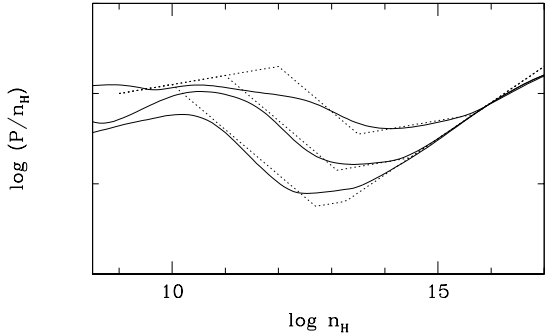}
\caption[dummy]{Effective equations of state (EOS) for clouds with 
$Z/Z_{\sun}=10^{-6}, 10^{-5}$ and $10^{-4}$ (solid). 
Also shown (dashed) are approximate relations (eq. \ref{eq:eos}) 
used in the hydrodynamical calculation.}
\label{fig1}
\end{figure}

Godunov-type smoothed particle hydrodynamics (SPH) with a second-order 
Riemann solver are used for hydrodynamics and 
a Tree method is used for gravitational force 
(Tsuribe \& Inutsuka 1999a).
The number of SPH particles is $N=1.6$ (in Runs A, B, D, F, and G) or 
$2.3\times 10^{6}$ (in Runs C and E), which is 
sufficiently large to meet the resolution condition for physical 
fragmentation (Klein et al 2004).

\section{Results}
Major parameters in this model are the metallicity $Z$, which 
determines the equation of state, and the initial amplitude of 
elongation ${\cal{E}}_0$.
The numerical results can be classified into two types 
according to whether the core fragments or not. 
In the following, we show the typical models for these cases.

\subsection{Cores with Fragmentation}
Figure 2 shows the evolution of the core with $Z=10^{-5}Z_{\odot}$ and 
${\cal{E}}_0=1.0$.
Initially, while the core is close to spherical, 
the evolution of elongation obeys the prediction by
the linear analysis of a $\gamma=1.1$ cloud. 
For central densities $n \ga n_{1} (=10^{11} {\rm cm}^{-3})$, 
the temperature decreases rapidly owing to the dust cooling, 
and the core collapses at almost the free-fall rate. 
In this phase, the elongation grows in proportion to  
$\rho^m$ with $m=0.354-0.5$, which is between the rate for spherical 
pressure-free collapse ($m=0.354$) and that for filament collapse ($m=0.5$).
In other words, the rate $m$ is enhanced from the linear value for a
sphere owing to a non-linear effect.
The equation of state begins to stiffen to $\gamma=1.1$ 
at $n=n_{2} (=10^{13}{\rm cm}^{-3})$, 
where the dust and gas couple thermally, 
and become adiabatic ($\gamma=1.4$) at 
$n=n_{3} (=3\times10^{14}{\rm cm}^{-3})$, 
where the core becomes optically thick. 
However, the core continues to elongate owing to inertia until 
$n \sim 10^{16}{\rm cm}^{-3}$, where the maximum value of elongation 
${\cal{E}}_{\rm max}=61$ is attained.
A long spindle forms in the high-density core.
Around this epoch, the spindle stops collapsing in a cylindrical 
radial direction and then fragments. 
Using the scale height of a filament 
$H=(2 c_s^2/\pi G \bar{\rho_c})^{1/2}$, where $c_s$ and $\bar{\rho_c}$ 
is isothermal sound speed and density averaged over the $z$-axis,  
the interval between each fragment is well described by $2 \pi H$,
corresponding to maximum growing mode in the linear stability 
analysis of equilibrium filaments 
(Nagasawa 1987; Inutsuka \& Miyama 1997).
The filament fragments into four pieces.
Since the fragments have an infall velocity along the $z$-axis, they merge 
with each other at the center, owing to small initial random velocities. 
This is partly because we do not include effects of rotation and 
angular momentum.
In spite of the merging of fragments at the center, 
the number of fragments does not decrease, at least 
within the present calculation, 
since new fragments continue to be born on the 
outskirts.
We also studied a case with the same parameters ($Z, {\cal{E}}_0$),
but with a different seed for the random number.
In this case, positions of density peaks were altered, but 
the distance between each fragment remained almost the same.
The dynamical evolution was also similar.

\begin{figure*}
\epsscale{0.75}  
\plotone{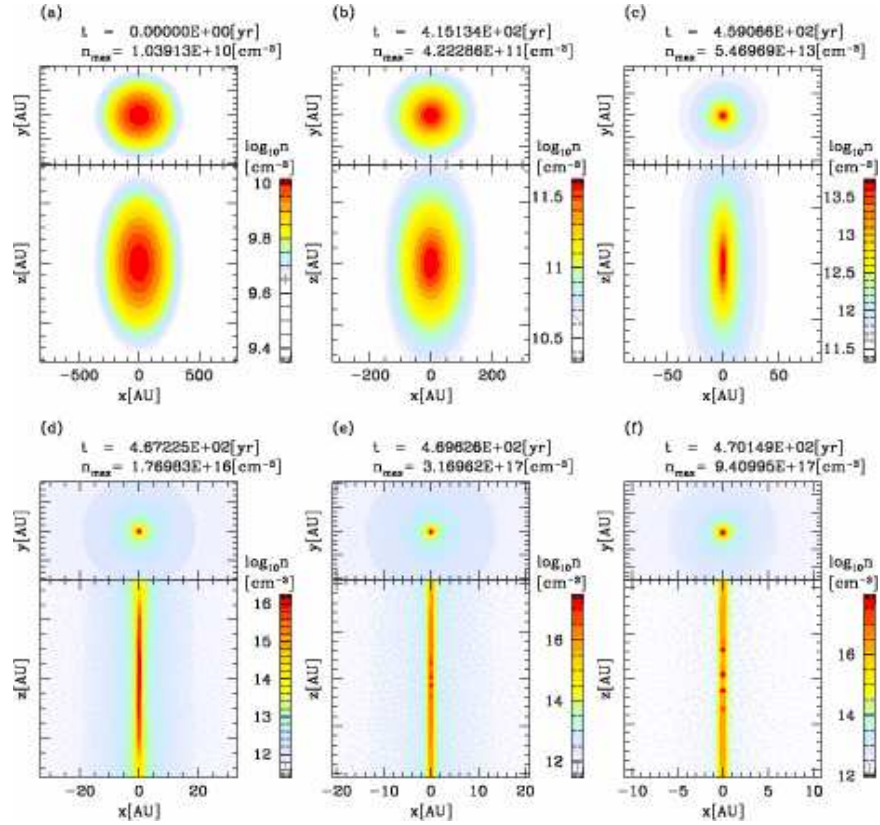}
\caption[dummy]{
Density distribution in the $x$-$y$ plane on the midplane 
(top panels), $x$-$z$ plane of $y=0$ (bottom panels)
for the case with ${\cal{E}}_0=1.0$ and 
$Z=10^{-5}Z_{\odot}$ at six different stages.
The color scale denotes the density in the logarithmic scale.
The maximum number density and time are shown in each panel.
}\label{fig2}
\end{figure*}
In the case of lower metallicity 
($Z=10^{-6}Z_{\odot}$, ${\cal{E}}_0=1.0$), 
the maximum elongation becomes smaller (${\cal{E}}_{\rm max} \sim 40$) 
because of a weaker effect of dust cooling.
Fragmentation is marginally observed. 
There are two fragments.

\subsection{Cores without Fragmentation} 
Models with smaller initial elongation exhibit different evolution.
As an example, we show one model with $Z=10^{-5}Z_{\odot}$ and 
${\cal{E}}_0=0.32$ in Figure 3. 
In the initial stage ($\gamma=1.1$), the evolution of elongation is 
very similar to examples with fragmentation (\S 3.1). 
Subsequently, in the cooling phase 
($n \ga n_{1}=10^{11} {\rm cm}^{-3}$), the elongation grows.
The growth rate, however, remains close to the linear growth rate 
($m=0.354$) because of the small initial elongation.  
Then, the growth of elongation slows down at 
$n_{3} \simeq 3\times 10^{14} {\rm cm}^{-3}$, 
where the equation of state becomes adiabatic.
The maximum elongation reaches $14$ in this case, but this is not 
enough to cause fragmentation.
Thereafter, the core becomes more spherical with contraction, 
instead of becoming cylindrical.
We also followed the evolution in the subsequent accretion stage 
by concentrating on the central region. 
No fragmentation was observed in our calculation.

\begin{figure*}
\epsscale{0.75} 
\plotone{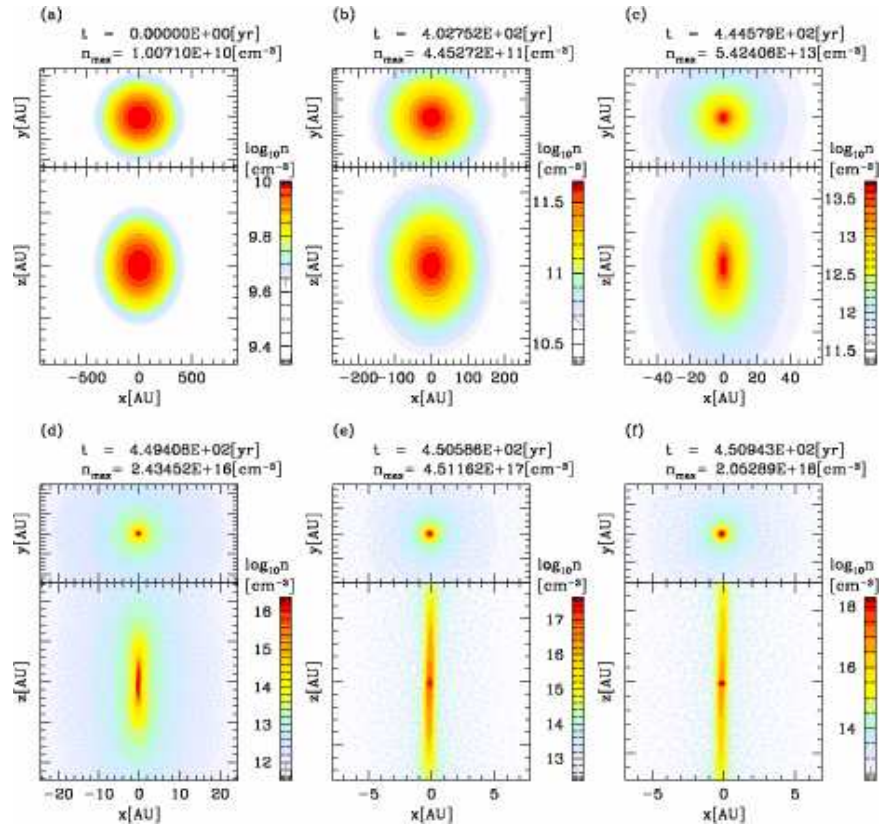}
\caption[dummy]{
Same as Fig. 1 but for the case with ${\cal{E}}_0=0.32$ and
$Z=10^{-5}Z_{\odot}$.
}\label{fig3}
\end{figure*}

In this calculation, the maximum value of axis ratio reaches 14.
This exceeds both the critical value for fragmentation $\pi$ and 
maximum growing mode $2\pi$ for fragmentation in the linear analysis
for uniform infinite filaments. 
Despite this, the filament does not fragment because of 
the central density concentration in our case.
Accretion onto the central density peak is faster than fragmentation
on the outskirts.  
In the cases with ${\cal{E}}_0=0.1$, no fragmentation is observed 
for any of the metallicities.

\section{Conclusion and Discussion}
We have studied the fragmentation process of low-metallicity 
($10^{-6}-10^{-4}Z_{\sun}$) cloud cores at high densities 
($\ga 10^{10}{\rm cm^{-3}}$).
Cores with initial axis ratios greater than about two become 
elongated enough to fragment via dust-cooling, even 
with metallicities of $10^{-6}-10^{-5}Z_{\sun}$.
This provides a possible mechanism to produce low-mass stars 
in ultra-metal-deficient environments.
The results of our calculations are summarized in Table 1. 
As a rule of thumb, fragmentation takes place if the core elongation
reaches a critical value ${\cal{E}}_{\rm crit} \ga 30$.
According to linear analysis for a filament of infinite length, namely,
without density gradient along the long axis, an equilibrium filament is
gravitationally unstable and subject to fragmentation if the axis ratio 
is greater than $\pi$ (Nagasawa 1987).
In our case, the filament is of finite length.
The density gradient toward the center along the long axis makes
the critical value for fragmentation several times larger than that for
a filament of infinite length as discussed in \S 3.2.
In Table 1, the maximum values of elongation in the linear analysis 
${\cal{E}}_{\rm max, L}$ are also presented.
We see that if ${\cal{E}}_{\rm max, L}$ exceeds  $\sim 3$, 
the elongation of the core will eventually reach 
${\cal{E}}_{\rm crit} \ga 30$ and the core fragments.
It should be noted that the actual maximum value of elongation and that
in linear theory are not reached at the same time because of non-linear
effects.  
The linear maximum value of elongation 
${\cal{E}}_{\rm max, L}$ is attained at the end of the rapid 
dust-cooling phase (i.e., $n=n_{2}$), since the elongation 
grows (or decays) if $\gamma < 1.09$ ($\gamma > 1.09$, respectively), 
in the linear theory (Hanawa \& Matsumoto 2000; Lai 2000).
On the other hand, the actual maximum is reached slightly after the core
becomes adiabatic (i.e., $n \ga n_{3}$).

Although we have demonstrated by detailed numerical work 
that dust-induced fragmentation indeed occurs under some circumstances, 
some points still remain to be improved.
To avoid too large a dynamic range, 
we were forced to take the initial state of our calculation 
at a rather high density of $10^{10}{\rm cm^{-3}}$ and 
parameterize the elongation of cores at this point.
The question of how frequently such an initial condition is realized 
remains uncertain.
The evolution after fragmentation is also to be studied. 
The final mass of stars is determined as a result of complex dynamical 
evolution such as merging and accretion.
In this phase, angular momentum would play an important role, 
although we have neglected effects of rotation for simplicity.  
For example, a disk may fragment into multiples during
the accretion phase (e.g., Saigo, Matsumoto, \& Umemura 2004). 
These issues should be investigated in a future work.

\acknowledgments
Numerical computations were carried out on VPP5000 at the Astronomical 
Data Analysis Center of the National Astronomical Observatory of Japan 
and in the computational facilities at Osaka University.
This research was supported in part by Grants-in-Aid for Young 
Scientists (B) 14740129 (TT) by the Ministry of Education, Culture, 
Sports, Science, and Technology of Japan (MEXT).



%
\begin{deluxetable*}{cccccccc}
\tablecaption{Model Parameters and Results
\label{tab:model} }
\tablehead{
\colhead{Run} &
\colhead{$\log{Z/Z_{\odot}}$} & 
\colhead{${\cal{E}}_0$} & 
\colhead{${\cal{E}}_{\rm max}$} & 
\colhead{$\log n_{\rm max}$} & 
\colhead{${\cal{E}}_{\rm max, L}$} & 
\colhead{Fragmentation} & 
} 
\startdata
A & -6 & 0.56 & 16   & 16.8 & 1.5 & No \\
B & -6 & 1.0  & 30   & 16.6 & 2.7 & Yes \\
C & -5 & 0.32 & 14   & 16.4 & 1.6 & No \\
D & -5 & 0.56 & 27   & 16.1 & 2.8 & Yes \\
E & -5 & 1.0  & 61   & 16.0 & 4.9 & Yes \\
F & -4 & 0.32 & 34   & 15.6 & 2.9 & Yes \\
G & -4 & 0.56 & 56   & 15.3 & 5.1 & Yes \\
\enddata
\tablenotetext{~}{NOTE: Model parameters are metallicity 
$\log{Z/Z_{\odot}}$ and the initial elongation ${\cal{E}}_0$. 
${\cal{E}}_{\rm max}$ is the maximum value of elongation, which is
attained at the number density $n_{\rm max}$. ${\cal{E}}_{\rm max, L}$ 
is the maximum value of elongation in the linear theory. }
\end{deluxetable*}

\end{document}